\documentclass[graybox, natbib, nosecnum]{svmult}
\bibpunct{(}{)}{;}{a}{}{,} % suppress commas between author-names and year

\usepackage{mathptmx}       % selects Times Roman as basic font
\usepackage{helvet}         % selects Helvetica as sans-serif font
\usepackage{courier}        % selects Courier as typewriter font
\usepackage{type1cm}        % activate if the above 3 fonts are
\usepackage{makeidx}         % allows index generation
\usepackage{graphicx}        % standard LaTeX graphics tool
\usepackage{multicol}        % used for the two-column index
\usepackage[bottom]{footmisc}% places footnotes at page bottom
\usepackage[normalem]{ulem}	% for strike-through of text with \sout{}  
\usepackage{hyperref}  %for hyperlinks
\usepackage{soul}   % for high-lighting of text

\begin{document}

\title*{Real-Time Simulation in Real-Time Systems: Current Status, Research Challenges and A Way Forward}
\author{Xi Zheng}
\institute{Xi Zheng \at Macquarie University, Macquarie University, NSW 2109, Australia \email{james.zheng@mq.edu.au}
%\and Name of Second Author \at Name, Address of Institute \email{name@email.address}
}
\maketitle
\abstract{Simulation especially real-time simulation have been widely used for the design and testing of real-time systems. The advancement of simulation tools has largely attributed to the evolution of computing technologies.   With the reduced cost and dramatically improved performance,  researchers and industry engineers are able to access variety of effective and highly performing simulation tools. This chapter describes the definition and importance of real-time simulation for real-time systems.  Moreover, the chapter also points out the challenges met in real-time simulation and walks through some promising research progress in addressing some of the challenges\footnote{the manuscript is submitted to Real-Time systems section of Handbook of Real-Time Computing published by Springer}.}

\section{Introduction}
Real-time systems, especially nowadays Cyber-Physical Systems (CPS)~\cite{baheti2011cyber}, entail digital devices which interact closely with analog ones, humans, and the surround world.  These systems are increasingly used in our daily lives including autonomous vehicles, smart grids, automated health care, and many other life-critical applications~\cite{zheng2014state,eriksen2001seaglider, zheng2015verification,wang2017learning,zheng2017perceptions}.  These real-time systems need to be reliable and secure and meet strict compliance check.   There have been many research efforts to improve and ensure the reliability and security of such systems.  For instance, formal methods~\cite{zheng2018efficient,cassez2017refinement,zheng2015braceassertion,bouyer2011timed,zheng2014physically, 
zeng2018aua,xu2019csp,pan2017cyber,radhappa2018practical,zheng2017security} and testing have been widely used to improve the quality in real-time systems.  However, the tigher coupling between physical processes and software components in the modern real-time systems along with the varying spatial and temporal runtime contexts made such system exhibit diverse behaviours across runs~\cite{zheng2017perceptions}.  Thus, verifying and validating real-time systems are still complicated where validation assures that a real-time system meets the needs of the customers and verification assesses whether a real-time system complies with the given specifciation~\cite{zheng2015verification} .

In industry, simulators have been used extensively in the planning, designing, implementation, and verification stage of real-time systems for decades~\cite{RealTimeSimulation}. With the above-mentioned complexity increased dramatically for modern real-time systems, the industry has a strong tendency of relying more heavily on simulaton tools~\cite{yamane2016real}. Nowadays the industry mainly uses real-time simulators to design and test various aspects of real-time systems including controls and protection schemes and devices.  The simulators allow the designer to conduct a variety of test earlier at the planning and design stage and in a repetitive and safe fashion. At the same time, with the rapid development of computing technologies, the cost of running simulations is steadily reduced and the performance is increasingly improved, making the simulators available to more researchers and engineers for a wider variety of real-time systems~\cite{zheng2015physically}.  The adoption of real-time simulation allows testing real-time sytems under faulty and extreme conditions without damaging equipments under test while maintaining sufficient flexibility in choosing test parameters and components.   Other benefits of using real-time simulations also include maintaining a relatively safe testing environment for the engineers and other personnel~\cite{zheng2017perceptions,nguyen2017real}.  

Though real-time simulation allows a tight coupling between a real hardware with a simulation tool to test  hardware or software components under realistic conditions,  the executation of the simulator in this case requires each time step execution to meet the real-time constraints of the correponding phsical target modelled.  The costs of developing and running such simulators can be very prohibitive for a relatively complex real-time systems (e.g., autonomous vehicles)~\cite{zheng2017perceptions}. A recent study~\cite{zheng2017perceptions} finds out that the truthfulness of the real-time systems behavior in simulation-based approaches is often uncertain, and in practice the uncertainty and random disturbances in the real-time sytems cannot be coped with by even extensive simulation.  As a result, simulation-based approach often failes to verify mission-critical real-time systems and identify key failure points.    Also in the same report,  the high-learning curve associated with creating the models and scalability remain major issues for simulation-based approaches. 

The objective of this chaper is to provide an introduction of simulation and more relevant give some definitions of real-time simulation.  Then the chapter will explain the history of real-time simulation, which is followed by
explaining real-time simulation support for real-time systems.  Then the chaper will walk through the challenges in establishing accurate real-time simulation and some industry best practices.  Finally, this paper concludes with the discussion of some promising research directions in exploring real-time simulation for more robust and accurate real-time systems. 

\section{What is Simulation and Real-Time Simulation}
Simulation uses the operation of one system to represent the operation of another. Since digital computers 
are not able to record a continuous snapshot of a transient phenomena,  but rather recording a sequence of snapshots at discrete intervals, as a result, the simulation discussed in the paper is also known as digial or numerical simulation with discrete time step.  In such simulation, time usually moves forward in steps which are either of equal duration~\cite{dommel1969digital} or variable duration~\cite{sanchez1995variable}.  At each time step, some form of mathematical functions or equations (e.g., differential equations) are solved.  While most of the linear systems can be simulated using fixed time step,  variable time step simulation is more suitable for non-linear and high frequency dynamic systems to give the capabilities to study both fast and slow phenomena reflecting the observed system behavior~\cite{sanchez1995variable}.   For instance,  in power grid systems, analysis of such systems must cater for the voltage instability and unpredictable disturbances which can span hundreds or thousands of seconds.  Using fixed time step simulation with small steps to solve equations and integrations in such system is not efficient.  Conversely,  using fixed time step simulation with large time steps fails to capture the possible fast transients associated with above-mentioned instability and disturbance.  As a result, a simulation is required to adjust automatically the time step~\cite{sanchez1995variable}.

In discrete time simulation, the amount of real time required to solve the underlying mathematical equations, which represents the system at a specific time step,  is known as the execution time $T_ {E}$. However, the specified step size $T_{S}$ in most of time is either shorter or longer than $T_ {E}$.  In the above-mentioned simulation scenarios, which are also known as "offline" simulation,  the difference between $T_{S}$  and $T_ {E}$ is irrelevant as the main objective of such simulation is to obtain simulation results as soon as possible.  $T_ {E}$ can be dedendent on many factors, the main contributors are the computation power of the host machine and complexity of the mathmatical equations (e.g., the system model) required to be solved in each time step.

In comparison, real-time simulation is an "online" version of discrete-time simulation, where time moves forward in steps of pre-defined duration~\cite{sanchez1995variable}. In real-time simulation, the simulation results are dependent not only on the equations/models but also on the $T_ {E}$~\cite{RealTimeSimulation}.
To solve the underlying mathematical equations (e.g., differential equations) at a specific time step, the model is solved using the input of the variables or states from the preceding time-step.  In each time-step, the real-time simulator is required to execute the same batch of tasks which include 1) reading inputs from the last time-step and generating outputs for the next time-step;  2) solving the equations (e.g., differential equations) specified for each time step; 3) if necessary,  exchanging data with other simulation nodes; 4) waiting (for the next time-step to start).  This implies that for any externally connected devices or simulation nodes,  the state of the simulated system can be exchanged only once at the begining of each time-step in the real-time simulation. 
This means, as compared to an ``offline'' version, the execution time $T_ {E}$ required to solve the equations for a time step must be shorter than the specified step size $T_{S}$.

\begin{figure}
    \centering
    \includegraphics{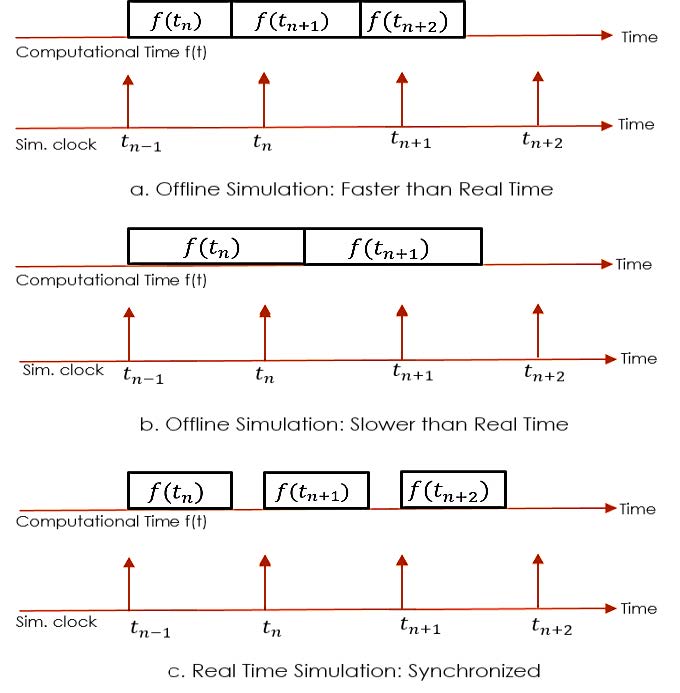}
    \caption{a. Accelerated Simulation b. Delayed Simulation c. Real-time Simulation~\cite{noureen2017real}}
    \label{fig:realtimesimulation}
    \vspace{0cm}
\end{figure}

To be more precise, for a real-time simulation to be valid, the simulator must calculate the values for those variables and states within the same time duration that the physical counterpart would require.  For instance, if it takes $10$ minutes to fill in a physical water tank, then the corresponding real-time simulator shall use a value of  $T_ {E}$, which is within $10$ minutes, to fill in a simulated water tank. 
Otherwise the real-time simulation is considered erroneous as discrepancies between the real-time simulation and its physical counterpart's responses are observed. This kind of error is commonly known as an "overrun"~\cite{RealTimeSimulation}. In Figure~\ref{fig:realtimesimulation}, (a) and (b) are examples of offline simulation where the actual execution time for the required mathematical equations is either shorter (a) (Accelerated Simulation) or longer (b) (Delayed Simulation) than a given simulation time step. The required time to solve the equations is largely dependent on the underlying mathematical function and corresponding variables.  In comparision, (c) depicts the real-time simulation scenario where the underlying mathematical function (e.g., $f(t_n)$) has to be done within the given discrete time-step (e.g., $t_n - t_{n-1}$) to avoid the "overrun". 
With the given discrete time-step, not only those equations inside the given function $f(t_n)$ need to be solved, but also the input variables of the function need to be processed,  and the results need to be output within and outside  the current simultion unit (e.g., current real-time simulation unit can interact with other real-time simulation unit or real hardware). 

%
%Since CPS often entail multiple physical processes, and each physical process may be modeled separately, different real-time simulations must be able to coordinate, even potentially exchanging state information during a single time step $T_{S}$.  Different simulation models and platforms may have different time steps, depending on their physical laws (e.g., a dynamic electrical system has a fast time step while a dynamic thermal system may have a much slower one).  Each real-time simulator (i.e., the executable implementing the simulation for a given model) has to execute a number of tasks within $T_{E}$, including reading inputs, solving model equations, generating outputs, and exchanging results with other simulation models.  All these tasks are important, and failures or inaccuracies in any of them can render the real-time simulation useless~\cite{RealTimeSimulation}.  Accurate synchronization among different simulation models is crucial to ensuring simulation stability~\cite{bednar2007stability}.

%{\bf Footnotes should not be used}!
\section{Evolution of Real-Time Simulation}

The state of the art real-time simulation can find its origin in physical simulation (i.e., analogue simulation). Early work in physical simulation in 1960's utilises  amplifiers, resistors, capacitors, diodes to simulate specific components of a physical system. This analogue simulation where physical components is physically connected to ech other in a manner similar to the real system is the basis for the Transient Network Analyses (TNA) and the High Voltage Direct Current (HVDC) simulators~\cite{RealTimeSimulation, kuffel1995rtds}. The advantage of this kind of simulation establishes a near one-to-one correlation between physical simulators and physical system components.  The advantage allows system engineers to design the simulation by mapping with the parameters/components of the target physical systems with minimum efforts spent on mapping and translation~\cite{brennan1964survey}.   However, users for physical simulation often struggle to solve those issues related to the computer rather than the simulation as a technique.  For instance, users for physical simulation need to scale all the equations (e.g., differential equations) and shall grasp the necessary and often advanced knowledge of the transformation solution (e.g., transforming those equations from the problem domain into the physical component of the computer).  

Digital simulations use mathematical representation to simulate the target physcial system.  The algorithms created for such software based simulations were described as early as in 1969 and have been used in some quite well known programs~\cite{kuffel1995rtds}. Compared with their analogue simulators which are able to operate in real time, these early digital simulators operate, however, in non real-time fashion.  As implied before, real-time operation requires that an event in the physical system which takes for one second shall be simulated on the corresponding simulator within one second.  Unfortunately in the early digital simulators , those computations necessary to solve the required mathematical equations might take many seconds or even minutes rather than within one second to meet the real-time simulation requirement.  With the quick development of microprocess, digital signal processing (DSP), together with the improvement of software modelling techniques,  digital real-time simulators started to replace its physical counterparts in 1980s~\cite{RealTimeSimulation}.   

The earlier pioneers in digital real-time simulation use relatively high-speed computers to achieve real-time operation for short period of time, but these simulators are restricted to quite simple systems~\cite{marti1994real,kezunovic1994transients}.  The main issue of these simulators lies in the scalability.  When the modelled system becomes realistic and more complex, the number of equations to be solved gets more complex.  Correspondingly, the number of computations needs to be performed restricts the applicability of these simulators.  To achieve real-time operations for digital simulators, many digital signal processors can work in parallel to share the computational tasks necessary to solve the underlying equations in the simulation models.  This DSP-based real-time simulators are developed with proprietary hardware and became commercially available~\cite{kuffel1995rtds}.  Since proprietary hardware imposes non-trivial limitation on its applicability,  some other digital real-time simulators based on commericial supercomputers are created (e.g., HYPERSIM from Hydro-Quebec~\cite{do1999hypersim}).  There are also other attempts to use low-cost standard PCs to host real-time simulators~\cite{hollman2003real}, and this attempt has been further accelerated by the introducing of low-cost Commerical Off-The-Shelf (COTS) multi-core processors.  COTS enabled digital simulators are able to conduct complex parallel simulation by reducing dependence on inter-computer communcations and these simulators are widely used to simulate large-scale microgridas, aircarft and power systems~\cite{belanger2007emegasim}.

A recent advancement in real-time simulation lies in running simulation models directly on Field-Programmable Gate Arrays (FPGAs).  This relatively new trend allows fully utilizing the parallel nature of FPGAs so that the time-step for the real-time simulation can be set to very small and a complex system can be simulated by many smaller models~\cite{saad2015real,chen2009fpga}.

\section{Real-Time Simulation Support for Real-Time Systems}

Real-Time Systems, especially recent Cyber-Physical Systems,  entail complex software and exhibit sophisticated interactions among digital devices, analog components, and the surrounding world, including humans in that world.  Such systems are often safety critical and must be reliable.  However, Real-Time Systems contain both digital and analog components and must be modeled as hybrid systems, which are known to be hard to formally verify~\cite{henzinger1995s}.  The state-of-the-practice in creating a repetitive and flexible test environment is to use real-time simulation, where computer models are used to accurately produce values of internal variables inside a real-time systems; these models are designed to operate on the same time-scale as the corresponding physical system~\cite{RealTimeSimulation}.   The basic assumption to use real-time simulation is to consider in a real-time system, the process which needs to be verified composes of a plant with a controller acting upon it. 
Thus, though real-time simulation have been used in various real-time systems,  these applications can be categorized as two tests: hardware-in-the-loop tests (HiL) ~\cite{chensimulink,zhang2013co} and software-in-the-loop tests (SiL) \cite{kwon1999real}. 

In HiL, a physical controller is connected to an executing real-time simulation representing a virtual plant, and this is used  to verify the controller.  Aircraft manufacturer Embraer used real-time simulation software platform  (e.g., RT-LAB)  to execute a highly accurate fighter plane model, which is connected with a real onboard aircraft computer and a real cockpit.  The real-time simulation model can provide a variety of feedback including force (via flight control joystick), visual and sound~\cite{de2004evolution}.   Industrial electronic company Mitsubishi also used real-time simulation to design motor drives, where a physical motor was simulated to work with its related real-world controllers~\cite{harakawa2005real}.  This simulation allows testing and verification of the whole system in a much earlier stage where a phsycal motor is not yet availalble for test.   Real-time simulation also helps to understand the integration of microgrid devices with renewable energy resources (e.g., solar power and wind farms), where the overall stability and transient responses from the integrated power system can be thorougly studied and various statiscal studies can be conducted for optimization and worse-case scenario analysis~\cite{paquin2007real,paquin2008hardware}.   

In comparision,  in SiL~\cite{kwon1999real}, both controller and plant are simulated.  SiL supports the embedded software in a real-time system to be tested  as early as possible where the entire real-world platform including equipment (which might not be ready yet) and environment (which can not be thoroughly tested) is modeled in software and simulated.   Compared with HiL which is often used during the testing phase of a real-time system, SiL can be used at all stages in a real-time system including design, development and testing~\cite{demers2007generic}. In~\cite{bayha2012model}, SiL is used to model the controllers and the enviornment of a unmanned aerial vehicles (UAV) system and test the control software using a variety of test cases for fault-tolerance and robustness.  In~\cite{demers2007generic}, SiL is used to evaluate a policy-based network management software against a variety of network simualtors.  In~\cite{muresan2012software}, SiL is used to evaluate the controller software for an electric motor and it was found the cost for implementing a SiL enviornment are about sixty times cheaper than HiL environment. Similarly in~\cite{russo2007software}, an SiL analysis is performed to evaluate a brake controller algorithm and good results are obtained by SiL with a cost effective way.

Though SiL is cost effective to produce repeatable results (as the randomness of the controller and plant both reside in the real-time simulator) and supports some basic forms fo testing (e.g., by visualizing the internal state of some variables), advanced debugging which allows tracabilities to software errors is not available and SiL does not provide any form of automation of test cases and test oracles neither~\cite{demers2007generic,bayha2012model}.

\section{Challenges and Best Practices in Industry}
Since real-time systems often entail multiple physical processes, and each physical process may be modeled separately, in Industry, different real-time simulations must be able to coordinate, even potentially exchanging state information during a single time step $T_{S}$.  Different simulation models and platforms may have different time steps, depending on their physical laws (e.g., a dynamic electrical system has a fast time step while a dynamic thermal system may have a much slower one).  Each real-time simulator (i.e., the executable implementing the simulation for a given model) has to execute a number of tasks within $T_{E}$, including reading inputs, solving model equations, generating outputs, and exchanging results with other simulation models.  All these tasks are important, and failures or inaccuracies in any of them can render the real-time simulation useless~\cite{RealTimeSimulation}.  Accurate synchronization among different simulation models is crucial to ensuring simulation stability~\cite{bednar2007stability}.

The Functional Mockup Interface (FMI)~\cite{blochwitz2012functional} is an independent standard to create a co-simulation environment where C code for a specific dynamic system model is generated in the form of an input/output block, and two or more models (with different solvers) can be coupled.  FMI requires each simulation platform provider (where each dynamic model is created) to explicitly support an FMI interface for model exchange so that it is possible to automatically generate a Functional Markup Unit (FMU) from the dynamic model. A FMU is a combination of C code and a helper XML specification that has definitions for all the variables in the given dynamic model. However, the two fundamental challenges in establishing real-time simulation, namely time synchronization and data integration among simulation models are left for developers to implement in the form of Master Algorithm. The MODELISAR~\cite{Modelisar} project supports FMI and includes a prototypical implementation of a Master Algorithm. However, the existing implementation does not guarantee the efficiency and simulation speed, which largely depend on the problem to be solved (e.g., the size of the underlying ordinary differential equation or differential algebraic equation) and the host computer's power~\cite{bastian2011master}. This kind of implementation of the Master Algorithm is not acceptable for an integrated  verification environment where efficiency and speed of the real-time simulation must be optimized to guarantee necessary precision of the outputs; further, writing a suitable master algorithm is very error-prone and poses significant challenges for developers~\cite{bastian2011master}.  Since numerical integrations deal with approximations, it is of vital importance to have an alternative automated solution that can guarantee efficiency and speed of the real-time simulation (instead of an interface or a requirement for data integration and time synchronization) to maintain a satisfactory balance between the simulation speed (i.e., latency) and precision (i.e., simulation errors)~\cite{khaled2014fast}.  In~\cite{al2012comprehensive}, a co-simulation platform is proposed to integrate the {\em ns-2} network simulator with the {\em Modelica} physical systems simulator. The simulation platform is able to support asynchronous events inside both physical and network systems. The main contribution of the work is to solve real-time synchronization to make sure both simulators will advance at the same wall-clock rate.  

In industry, real-time system practitioners guarantee the simulation speed and precision by using dedicated machines and software to build the real-time simulation environments (i.e., NI PXI server~\cite{PXI} and LabVIEW real-time module~\cite{LabViewRT}).  However, this approach is very expensive (e.g., a basic NI PXI server costs around $10,000$ USD~\cite{PXIPrice}) and is not scalable.  Also this approach does not provide a solution for complex real-time systems where sub-system models must be created in different simulation platforms for a variety of reasons (e.g., knowledge and preference of the interdisciplinary team, different costs, and different built-in solvers).  In~\cite{khaled2014fast}, a fine-grained co-simulation method is explored that enables numerical integration speed-ups. The method is to partition the existing models into loosely coupled sub-systems with sparse communication between partitioned modules. The parallel execution is mainly to exploit multi-core processors to deal with originally sequential ordinary differential equations in real-time system's sub-system models.  In~\cite{kinsy2011time}, a time-predictable computer architecture for digital emulation is proposed for real-time systems. The architecture can be implemented on top of a Field Programmable Gate Array (FPGA) to provide low latency emulation.  In~\cite{yan2012integrated}, an integrated platform is proposed to integrate Matlab/Simulink simulation tool with the DETERLab emulation testbed.  The runtime environment provides time synchronization and data communication to coordinate two simulation platforms for security experiments.

%However, to date, the application of real-time simulation in CPS verification is limited. 
%Besides being expensive (e.g., requiring proprietary hardware) and bound to a specific simulation platform (e.g., Simulink), all these approaches use testing to check the controller algorithm that  resides in the cyber components of a CPS.  In comparison, an
%  intuitive and generic way of leveraging real-time simulation with a
%  more formal runtime verification would give more thorough bug
%  detection not limited by the coverage of a particular suite of
%  tests.
%
%Simulation based approaches to verification, on the other hand, are restrictive both in expressiveness (e.g., of quantitative properties) and coverage (i.e., the cyber part is {\em modeled} instead of testing
%the real implementation). Thus common but subtle bugs that result from the interaction of cyber and physical components are often not detectable.

\section{Future Direction}
Simulation based approaches are widely used in industry-scale real time systems, however, they are restrictive both in expressiveness (e.g., of quantitative properties) and coverage (i.e., the cyber part is modeled instead
of testing the real implementation). Thus common but subtle bugs that result from the interaction
of cyber and physical components in complex real-time systmes (e.g., Autonomous vehicles) are often not detectable. 

On the other hand, runtime verification where properties are formally specified and checked at runtime, receives a lot of attention to verify real-time systems.  In~\cite{BraceAssertion2015}, runtime monitors can check both qualitative (e.g., safety, liveness) and quantitative (e.g., bounded safety and liveness, responsiveness) properties.  However, to detect the insidious real-time systems bugs that are manifest only in a specific deployment environment,  runtime verification techniques require repetitive deployments that are either too expensive (e.g., in labor, time, and/or money), dangerous (e.g., involving autonomous vehicles), or infeasible.  As an example, an unmanned rover deployed to the moon was unable to move after the first lunar night.  A {\em post hoc} analysis found that the temperature on the moon is considerably lower than the rover's components had accounted for; as a result, the rover effectively suffered from frostbite\footnote{Chen, Stephen. ``Last-ditch efforts to salvage mission of China's stricken Jane Rabbit Lunar rover.''
 {\it South China Morning Post} 18 April 2014, (\url{http://tinyurl.com/oq5qnqx})}.  Runtime verification of real-time systems in
general requires a repetitive and flexible test environment where settings can be changed easily to determine whether the properties being checked will hold in all situations. In the case of the rover, there are relatively accurate models of involved physical processes (e.g., rover dynamics and moon environment). However these models are separate from the runtime verification of the system's cyber components, ultimately leading to the failure. 

In~\cite{zheng2017real},  a combined verification approach that allows real-time system developers to opportunistically leverage real-time simulation to support runtime verification. The middleware, termed {\em BraceBind}, allows selecting, at runtime, between actual physical processes or simulations of them to support a running real-time system. {\em BraceBind} is  a real-time simulation architecture to generate and manage multiple real-time simulation environments based on existing simulation models in a manner that ensures sufficient accuracy for verifying a real-time system  Specifically, {\em BraceBind} aims to both improve simulation speed and minimize latency, thereby making it feasible to integrate simulations of physical processes into the running real-time system. {\em BraceBind} then integrates this real-time simulation architecture with an existing runtime verification approach that has low computational overhead and high accuracy. This integration uses an aspect-oriented adapter architecture that connects the variables in the cyber portion of the real-time system with either sensors and actuators in the physical world or the automatically generated real-time simulation. Their experimental results show that, with a negligible performance penalty, {\em BraceBind} is both efficient and effective in detecting program errors that are otherwise only detectable in a physical deployment.

Another promising direction to improve real-time simulation lies in the increasingly popular machine learning and deep learning models, which have been extensively in real-time systems to detect object location~\cite{pan2018object}, user activites~\cite{lu2019detection,bhandari2017non}, driver drowsiness~\cite{zhang2019driver}, road conditions for vehicles~\cite{zhou2019road,xie2018hybrid}.  As real-time systems are highly complex and essentially probalistic,  deep learning models can be used along with real-time simulators to improve the robustness and accuracy of simulation models. There are already some pioneering work towards this path as in ~\cite{vedaldi2015matconvnet,sivanandam2006introduction}.  It would be very exciting and interesting to see how convolutional neural network (CNN) and recurrent neural network (RNN)~\cite{haykin1994neural} can be used effectively with real-time simulation to improve robusness and safety of real-time systems.

\bibliographystyle{authordate1}
\bibliography{references}

\end{document}